\documentclass[9pt,twocolumn]{IEEEtran}
\usepackage[ruled, linesnumbered]{algorithm2e} 
\usepackage[utf8]{inputenc} 
\usepackage[T1]{fontenc} 
\usepackage{amsmath, amssymb, amsfonts} 
\usepackage{graphicx} 
\usepackage{subcaption} 
\usepackage[font=footnotesize]{caption} 
\usepackage{booktabs, array, tabularx, multirow} 
\usepackage{float} 
\usepackage{enumitem} 
\usepackage[colorlinks=true, allcolors=blue]{hyperref} 
\usepackage{graphicx} 
\usepackage{stfloats} 
\usepackage[margin=0.75in]{geometry} 
\usepackage[ruled, linesnumbered]{algorithm2e} 
\usepackage{amsmath, amssymb, amsfonts} 
\usepackage{cite} 
\usepackage{amsmath, amssymb, amsfonts}
\usepackage{balance}
\renewcommand{\baselinestretch}{0.952} 
\begin{document}
\newcommand\AtPageUpperMyright[1]{\AtPageUpperLeft{
 \put(\LenToUnit{0.28\paperwidth},\LenToUnit{-1cm}){
     \parbox{0.78\textwidth}{\raggedleft\fontsize{9}{11}\selectfont #1}}
 }}

\title{Optimizing Deep Learning for Skin Cancer Classification: A Computationally Efficient CNN with Minimal Accuracy Trade-Off \\}

\author{
\IEEEauthorblockN{
Abdullah Al Mamun\IEEEauthorrefmark{1},
Pollob Chandra Ray\IEEEauthorrefmark{1},
Md Rahat Ul Nasib\IEEEauthorrefmark{2},
Akash Das\IEEEauthorrefmark{3},
Jia Uddin\IEEEauthorrefmark{4},
Md Nurul Absur\IEEEauthorrefmark{5}
}\\
\IEEEauthorblockA{\IEEEauthorrefmark{1}Dhaka University of Engineering \& Technology, Bangladesh\\
\IEEEauthorrefmark{2}Samsung Austin Research Center, United States\\
\IEEEauthorrefmark{3}Federation University Australia, Australia\\
\IEEEauthorrefmark{4}Woosong University, Korea\\
\IEEEauthorrefmark{5}Graduate Center, CUNY, United States\\
\{mamun.duet.bd, pollob.cray, nasib131\}@gmail.com,
akashd@students.federation.edu.au,\\
jia.uddin@wsu.ac.kr, mabsur@gradcenter.cuny.edu}
}

\maketitle

\begin{abstract}
The rapid advancement of deep learning in medical image analysis has greatly enhanced the accuracy of skin cancer classification. However, current state-of-the-art models, especially those based on transfer learning like ResNet50, come with significant computational overhead, rendering them impractical for deployment in resource-constrained environments. This study proposes a custom CNN model that achieves a 96.7\% reduction in parameters (from 23.9 million in ResNet50 to 692,000) while maintaining a classification accuracy deviation of less than 0.022\%. Our empirical analysis of the HAM10000 dataset reveals that although transfer learning models provide a marginal accuracy improvement of approximately 0.022\%, they result in a staggering 13,216.76\% increase in FLOPs, considerably raising computational costs and inference latency. In contrast, our lightweight CNN architecture, which encompasses only 30.04 million FLOPs compared to ResNet50's 4.00 billion, significantly reduces energy consumption, memory footprint, and inference time. These findings underscore the trade-off between the complexity of deep models and their real-world feasibility, positioning our optimized CNN as a practical solution for mobile and edge-based skin cancer diagnostics.
\end{abstract}


\section{Introduction}

The emerging era in the medical domain and healthcare sector is marked by complex machine learning implementations and robust solutions to enhance accuracy and efficiency in the medical field and real-life patient monitoring \cite{Nova2021} \cite{10651815}. Advances in data management have introduced comprehensive medical datasets that require careful processing to ensure they are more translatable and actionable for timely disease identification. Innovative neural architectures, such as transformers and diffusion models, enable early diagnosis, personalized treatment, and predictive analytics with unprecedented accuracy \cite{10.1007/978-981-19-9483-8_32}. Given the rising fatality rates associated with skin cancer \cite{10430475}, there is an urgent need for a closer examination of the existing image datasets related to skin cancers. By applying advanced machine learning algorithms, we can improve early detection methods and advance the capabilities of skin cancer detection.

Among the various types of skin cancer, melanoma is regarded as the most dangerous due to its high level of aggressiveness \cite{cancers14194652}. Early melanoma detection significantly improves survival chances; however, recognizing the disease in its initial stages remains challenging. Traditional diagnostic methods primarily rely on clinical examinations and biopsy procedures, which can be time-consuming and invasive. As a result, automated systems for skin cancer detection, particularly for melanoma, have garnered considerable attention in the medical community. Deep learning—an advanced approach to automated image analysis and medical diagnostics—has been effectively utilized in multiple areas, including the identification of skin lesions, tumor classification, and disease progression prediction \cite{10.1007/978-981-16-6460-1_3} \cite{Nova2022}. This method demonstrates superior accuracy and offers more efficient healthcare solutions than traditional machine learning techniques.

Convolutional Neural Network (CNN) applications for skin cancer classification form a significant part of current research studies. In \cite{9523164}, test results showed that CNNs hold promise for differentiating skin malignancies from benign lesions yet struggled to detect minimal-stage melanomas. Designed a model based on CNN principles, which delivered 88\% correctness yet proved ineffectual when analyzing early-stage melanoma lesions.\cite{8945133}

The analysis of the HAM10000 database resulted in a success rate of 94.3\%, as reported in their study \cite{Tschandl2018-qh}. The detection of early-stage melanoma continues to pose a diagnostic challenge within medical science. While VGG16 was employed, it yielded unsatisfactory results in identifying early-stage melanoma \cite{10486410}. To address this, a hybrid CNN model was developed with Long-Short-Term Memory (LSTM) networks, achieving an F1 score of 93.4\%. However, this approach necessitated heightened computational resource utilization and fell below real-time processing standards \cite{9133532}. Although efficient skin cancer classification has been accomplished using ResNet50, improving specificity in detecting early melanomas remains pressing. Additionally, managing costs for allocating resources across many+ parameters can complicate disease detection approaches and render them less effective from a real-life cost perspective \cite{https://doi.org/10.1002/cpe.7542} .

Addressing computation costs through Floating Point Operations (FLOPs) is one of the most precise metrics for understanding the overall cost of deep learning implementations \cite{Chen_2023_CVPR}. In \cite{pmlr-v238-meng24a}, optimizing FLOPs enhances accuracy by 48\% compared to state-of-the-art methods. Placing greater emphasis on managing FLOPs can aid in balancing the trade-off between custom CNN implementations and any pre-trained models, alongside their detection accuracy.

\textit{Our proposed system, leveraging a custom CNN architecture alongside the pre-trained ResNet50 deep-learning model, is specifically designed for the early detection of melanoma. We aim to enhance the model's performance to achieve high accuracy while also considering the computational cost, using FLOP metrics to assess the potential trade-offs between cost and accuracy. This consideration significantly contributes to our approach, emphasizing the balance between accuracy and processing efficiency.}

The rest of the paper is organized as follows. Section~\ref{sec: related work} discusses the related work in this area. Section~\ref{sec:systemmodel} introduces our proposed methodology, and Section~\ref{sec:experimentresults} discusses performance evaluation. Finally, Section~\ref{sec:conclusions} concludes the paper by addressing future works.

\section{Related Study}
\label{sec: related work}
Skin cancer detection has become a significant focus in medical imaging research, with deep learning techniques significantly improving diagnostic accuracy. Over the years, numerous studies have examined using convolutional neural networks (CNNs), transfer learning, and hybrid architectures to classify effectively and segment skin lesions. These models have achieved notable success in distinguishing between malignant and benign cases by utilizing extensive dermoscopic image datasets, such as ISIC and HAM10000. Additionally, researchers have incorporated multimodal data, including clinical metadata and patient history, to enhance predictive performance. This section offers an overview of key contributions in this field, emphasizing advancements in model architectures, dataset implementation, and performance evaluation strategies.

Table \ref{tab:related_studies_comparison} reviews several studies focused on skin cancer detection utilizing deep learning models, detailing the model used, the dataset, accuracy, and the advantages of each method. Deep learning techniques, including CNNs, hybrid CNN-LSTM SVM models, and ResNet50, are examined and applied to datasets such as ISIC 2018 and HAM10000 and open-source images. The accuracy of these models ranges from 76.0\% to 97.4\%, while our proposed system has an accuracy of 87.08\%. We adapt our proposed system to incorporate a more lightweight model to enhance efficiency and ensure the model is suitable for real-time use.

\begin{table}[htbp]
    \caption{Comparison of Related Studies in Skin Cancer Detection}
    \centering
    \scriptsize 
    \renewcommand{\arraystretch}{1.4} 
    \setlength{\tabcolsep}{2pt} 
    \begin{tabular}{p{0.6cm} p{2.2cm} p{1.2cm} p{0.8cm} p{0.8cm} p{2.2cm}}
        \toprule
        \textbf{Study} & \textbf{Model} & \textbf{Dataset} & \textbf{Acc.} & \textbf{Param.} & \textbf{Properties} \\
         & & & & \textbf{Red.} &  \\
        \midrule
        \cite{milton2019automatedskinlesionclassification} & PNASNet-5-Large Model & ISIC 2018 & 76.0\% & No & Ensemble Learning\\
        \cite{10212670} & CNN-LSTM-SVM & HAM10000 & 88.24\% & No & Temporal learning \\
        \cite{https://doi.org/10.1002/ima.22750} & DenseNet77 & ISIC 2018 & 91.4\% & No & Encoder-Decoder \\
        \cite{10796230} & Xception & Opensource & 91.0\% & No & SHAP Consideration \\
        \cite{10449996} & VGG19 & HAM10000 & 97.4\% & No & YOLO \\
        \cite{10080355} & SqueezeNet & HAM10000 & 88.2\% & No & Comparisons Based \\
        \cite{10770743} & IRv2-CASA & HAM10000 & 94.6\% & No & CASA Block  \\
        \textbf{Ours} & \textbf{ResNet50 + TL} & \textbf{HAM10000} & \textbf{89.08\%} & \textbf{Yes} & \textbf{Early Parameter Reduction} \\
        \bottomrule
    \end{tabular}
    \label{tab:related_studies_comparison}
\end{table}

Skin cancer detection has made significant strides thanks to recent investigations into deep learning techniques. The evaluation of deep learning models, such as ResNet50, InceptionV3, and DenseNet121, has shown that deep convolutional neural networks (CNNs) hold remarkable promise for medical imaging tasks related to skin cancer detection \cite{10122533} \cite{9706441}. The fundamental architecture of CNNs while highlighting their potential applications in dermatology and their value for identifying medical skin conditions. Furthermore, a hybrid system that combines CNNs with LSTM networks has been proposed to enhance the accuracy of automated skin cancer detection \cite{9696522}. Researchers have assessed various deep-learning approaches for skin cancer diagnosis, discussing the advantages and limitations of each model subtype. Additionally, a novel multi-feature integration approach has emerged, enhancing the classification results of deep learning models for skin cancer detection \cite{9701350}.

The cost of creating and running a deep learning framework raises significant concerns regarding computational usage. Authors of \cite{qi2024analysis} analyze the energy consumption of nine different ML models based on FLOP metrics to highlight the significance of computation. The DDPG algorithm reduces a substantial amount of FLOP, ensuring the efficiency of deep learning algorithms \cite{li2024mcmc}.

\section{Proposed Methodology}
\label{sec:systemmodel}
\subsection{Baseline Formulation}
\begin{figure}[htbp]
    \centering
    \includegraphics[width=3.5in]{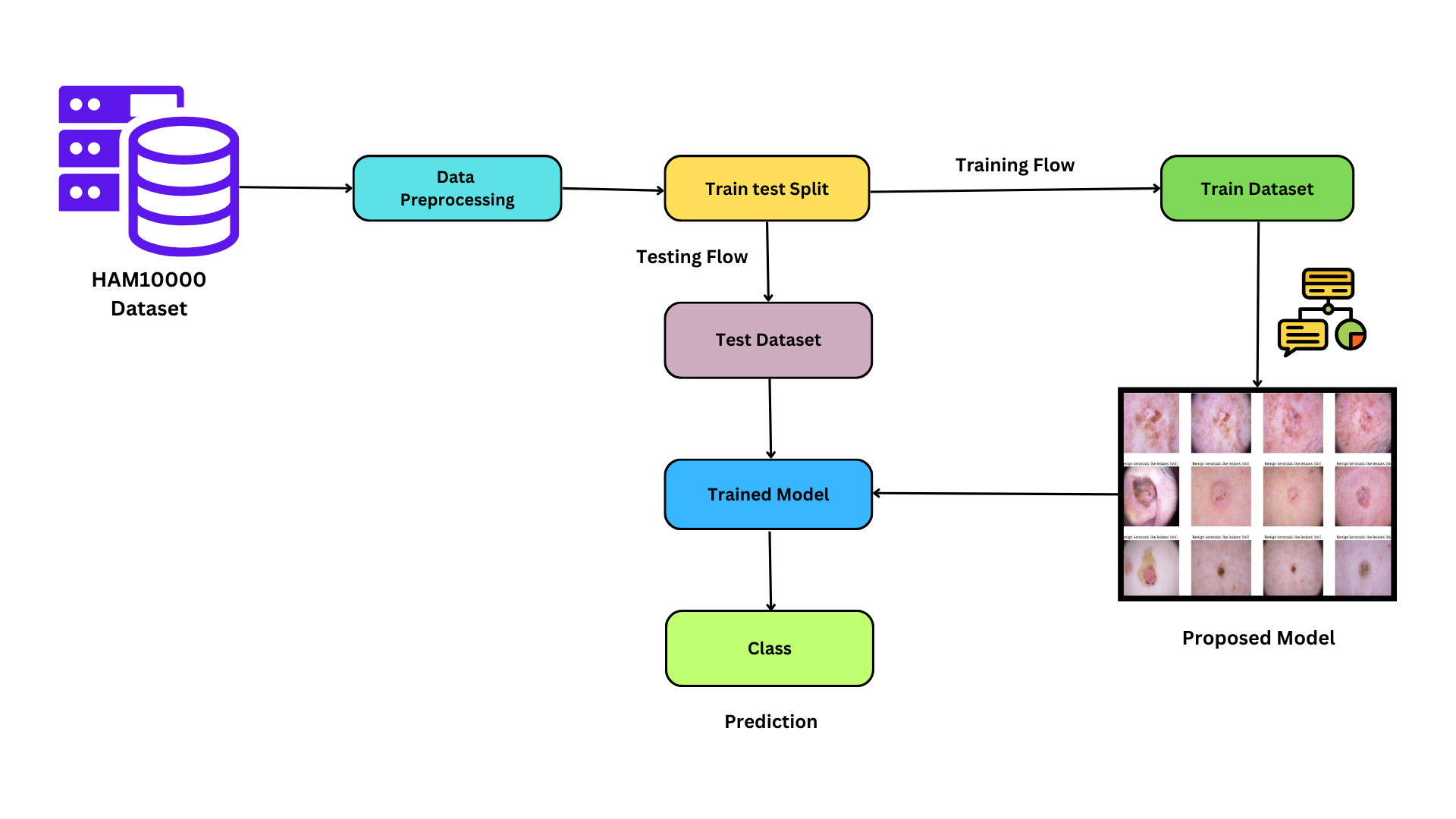}
    \caption{Proposed System Methodology}
    \label{fig:Figure-1}
\end{figure}

The proposed methodology (Figure - \ref{fig:Figure-1}) for skin detection employs the custom CNN \& lightweight ResNet50 model with transfer learning to minimize the number of parameters while preserving high classification accuracy and efficiency. Initially, the system gathers and preprocesses images from the HAM10000 dataset, resizing and normalizing them to ensure uniformity. At the heart of the system is the ResNet50 model, the baseline utilized for feature extraction by leveraging the pre-trained convolutional layers while omitting the fully connected layers. During the transfer learning phase, new fully connected layers are integrated into the network to fine-tune the model for the skin cancer detection task, with the pre-trained layers initially frozen to retain the learned features from the ImgNet dataset. The model is optimized using the Adam optimizer and the binary cross-entropy loss function, ensuring efficient convergence.

\begin{figure}[htbp]
    \centering
    \includegraphics[width=3.5in]{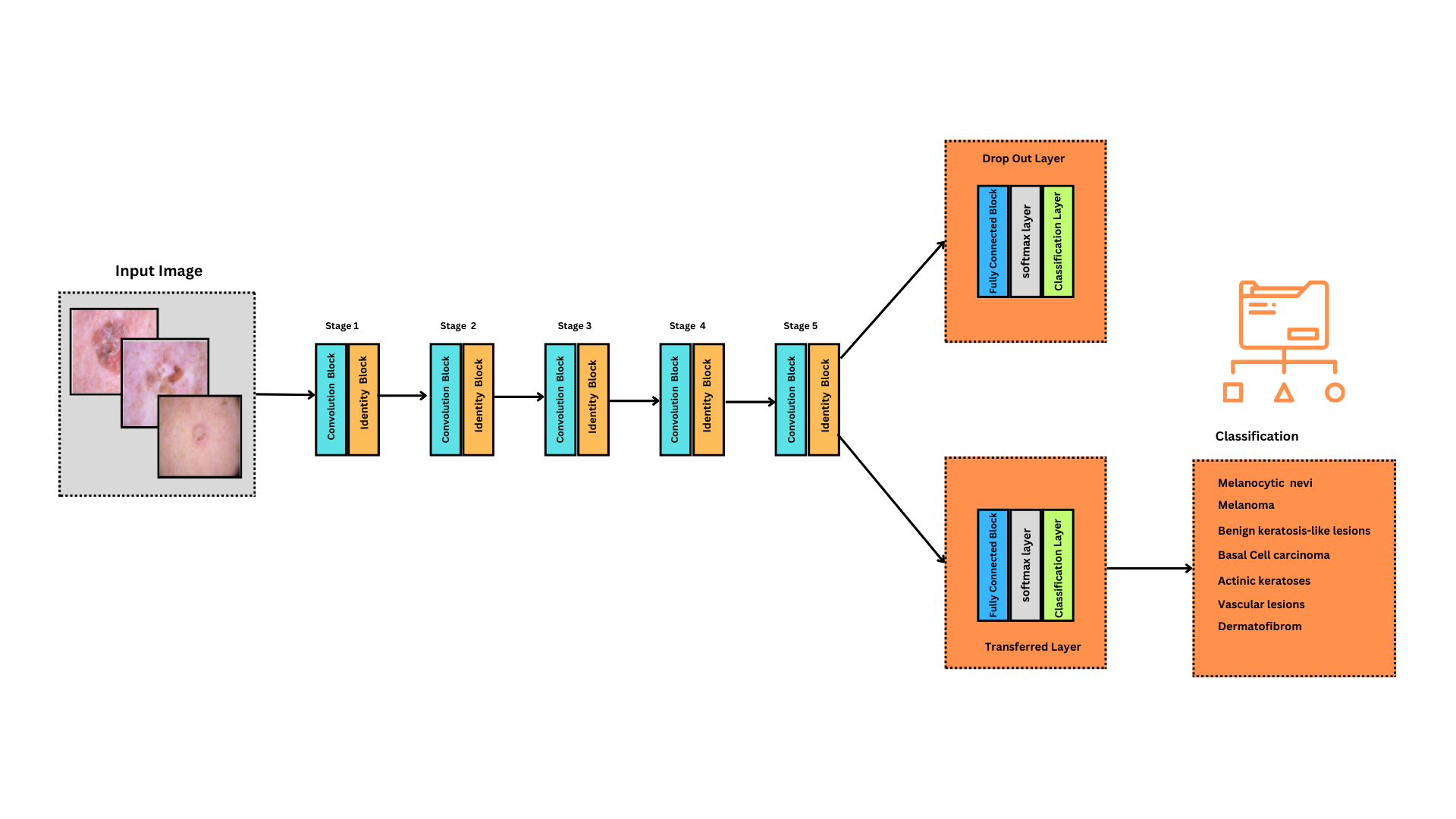}
    \caption{Proposed Model}
    \label{fig:Figure-2}
\end{figure}

Figure \ref{fig:Figure-2} and Table \ref{tab:custom_cnn_architecture} properly represent the proposed method and details of the proposed custom CNN architecture.

\vspace{-0.3cm}
\subsection{Data Collection and Preprocessing}
The skin cancer detection system utilizes data from the HAM10000 dataset \cite{DVN/DBW86T_2018}, one of the most comprehensive publicly accessible datasets designed explicitly for skin lesion analysis. This database contains over 10,000 diagnostic images representing both benign and malignant conditions. Before processing with the ResNet50 model, the images must be standardized to 224x224 pixel dimensions. Each input image undergoes normalization as a crucial part of preprocessing, ensuring pixel values are scaled between 0 and 1 for optimal model training. The normalization process can be mathematically expressed as follows:
\begin{equation}
    I_{\text{norm}} = \frac{I - \mu}{\sigma}
\end{equation}

Where:
\begin{itemize}
    \item \( I \) is the original pixel value of an image,
    \item \( \mu \) is the mean pixel value of the dataset,
    \item \( \sigma \) is the standard deviation of the pixel values of the dataset,
    \item \( I_{\text{norm}} \) is the normalized pixel value.
\end{itemize}

Additionally, data augmentation techniques such as random rotations, sequential correlations\cite{9303557}, and flips artificially enhance the diversity of the dataset. This enhancement strengthens the model’s robustness and generalization abilities. Exposing the model to a wide array of skin lesion appearances helps mitigate overfitting and improves its performance on unseen images.

\begin{table}[htbp]
\centering
\scriptsize
\caption{Layer-wise Details of the Proposed Custom CNN Architecture}
\begin{tabular}{lccccc}
\toprule
\textbf{Layer} & \textbf{Filters} & \textbf{Kernel} & \textbf{Act.} & \textbf{Output Shape} & \textbf{Params} \\
\midrule
Input         & -     & -     & -        & (224, 224, 3)   & 0 \\
Conv2D        & 32    & 3x3   & ReLU     & (224, 224, 32)  & 896 \\
MaxPool2D     & -     & 2x2   & -        & (112, 112, 32)  & 0 \\
Conv2D        & 64    & 3x3   & ReLU     & (112, 112, 64)  & 18,496 \\
MaxPool2D     & -     & 2x2   & -        & (56, 56, 64)    & 0 \\
Conv2D        & 128   & 3x3   & ReLU     & (56, 56, 128)   & 73,856 \\
MaxPool2D     & -     & 2x2   & -        & (28, 28, 128)   & 0 \\
Flatten       & -     & -     & -        & (100352,)       & 0 \\
Dense         & 256   & -     & ReLU     & (256,)          & 25,690,112 \\
Dropout (0.5) & -     & -     & -        & (256,)          & 0 \\
Dense (Out)   & 7     & -     & Softmax  & (7,)            & 1,799 \\
\bottomrule
\end{tabular}
\label{tab:custom_cnn_architecture}
\end{table}

\subsection{Feature Extraction with ResNet50}
Feature extraction is performed using the \textit{ResNet50} model, a deep convolutional neural network (CNN) designed to address the challenges posed by deep architectures. The ResNet50 model employs residual connections that allow the network to bypass specific layers, preventing the vanishing gradient problem and facilitating more efficient training for deeper models. The model begins by passing the input image \textit{I} through several convolutional layers, each applying filters to extract feature maps. The key feature of ResNet50 is the use of residual blocks, where the input to the block \( x \) is added to the output \( F(x) \), as defined by the following equation:

\begin{equation}
   y = F(x, \{W_i\}) + x 
\end{equation}

Where:

\begin{itemize}
    \item \( x \) is the input to the residual block,
    \item \( F(x, \{W_i\}) \) is the learned transformation (convolution, batch normalization, etc.),
    \item \( W_i \) are the weights of the block's filters,
    \item \( y \) is the output after adding the residual connection.
\end{itemize}
This residual learning reduces the degradation problem, allowing the model to attain improved accuracy with deeper layers. After the image passes through the residual blocks, the output feature maps are pooled and flattened to form a feature vector. This feature vector is subsequently input into fully connected layers, where it is converted into class probabilities using the softmax activation function:
\begin{equation}
P(y_i | x) = \frac{e^{z_i}}{\sum_{j} e^{z_j}}
\end{equation}

Where:
\begin{itemize}
    \item \( P(y_i | x) \) is the probability of class \( y_i \) given the input \( x \),
    \item \( z_i \) is the output of the final layer for class \( y_i \),
    \item The denominator \( \sum_{j} e^{z_j} \) sums over all classes \( j \),
\end{itemize}
The output pertains to various classes in skin cancer detection. ResNet50's pre-trained weights, originally trained on the ImageNet dataset, have been fine-tuned using the HAM10000 dataset to tailor the model specifically for skin cancer detection tasks. This methodology enables extracting hierarchical features, from basic textures to intricate patterns unique to skin lesions, thereby positioning ResNet50 as an excellent choice for automated skin cancer classification.

\subsection{Transfer Learning for Fine-Tuning}
After extracting features using ResNet50, transfer learning is conducted by introducing custom fully connected layers to the feature extractor for classification purposes. Initially, the pre-trained layers of ResNet50 are frozen, ensuring their weights remain unchanged during training. Subsequently, the newly added fully connected layers are trained to classify the images into benign or malignant categories. The loss function utilized during training is binary cross-entropy, a standard choice for binary classification tasks. The binary cross-entropy loss function is defined as follows:

The loss function used during training is binary cross-entropy, a widely adopted method for binary classification tasks. The binary cross-entropy loss function is given by:

\begin{equation}
L_{binary} = - \frac{1}{N} \sum_{i=1}^{N} \left[ y_i \log(p_i) + (1 - y_i) \log(1 - p_i) \right]
\end{equation}

Where:
\begin{itemize}
    \item \( N \) is the number of samples,
    \item \( y_i \) is the true label (0 for benign, 1 for malignant),
    \item \( p_i \) is the predicted probability of the sample being malignant.
\end{itemize}
\subsection{Fine-Tuning and Model Optimization}
After training the newly added layers, some of the earlier layers of ResNet50 are unfrozen for fine-tuning. This allows the model to make slight adjustments to the pre-trained weights and better adapt to the specific task of skin cancer detection. This fine-tuning process enhances performance by enabling the model to learn domain-specific features from the dataset while retaining the general features acquired during pre-training. We use the Adam optimizer to optimize the model during this phase, effectively combining the advantages of momentum and adaptive learning rates. The Adam update rule for the weights \( \theta \) is given by the following equation:

\begin{equation}
\theta_t = \theta_{t-1} - \eta \frac{v_t}{\sqrt{m_t} + \epsilon}
\end{equation}

Where:
\begin{itemize}
    \item \( \eta \) is the learning rate,
    \item \( m_t \) is the first moment estimate (momentum),
    \item \( v_t \) is the second moment estimate (adaptive learning rate),
    \item \( \epsilon \) is a small constant to avoid division by zero.
\end{itemize}
\subsection{Model Evaluation}
After training and fine-tuning the model, evaluating its performance is crucial to ensure its effectiveness in detecting skin cancer. The primary metric employed for this evaluation is accuracy, representing the proportion of correctly classified images out of the total images in the test set. Accuracy can be expressed as:

\begin{equation}
\text{Accuracy} = \frac{\text{Number of Correct Predictions}}{\text{Total Number of Predictions}}
\end{equation}

Alongside accuracy, we utilize a confusion matrix to calculate additional metrics like precision, recall, and F1 score, providing deeper insights into the model's performance. 

Precision is the ratio of true positive predictions (correctly predicted malignant instances) to all positive predictions (including true and false positives). It is defined as:

\begin{equation}
\text{Precision} = \frac{TP}{TP + FP}
\end{equation}

Where \( TP \) is the number of true positives, and \( FP \) is the number of false positives.

Recall is the ratio of true positive predictions to the total number of positive instances (true positives plus false negatives). It is defined as:

\begin{equation}
\text{Recall} = \frac{TP}{TP + FN}
\end{equation}

Where \( FN \) is the number of false negatives.

Finally, the F1 score is the harmonic mean of precision and recall, providing a balanced measure that considers both metrics. It is calculated as:

\begin{equation}
\text{F1-Score} = \frac{2 \cdot \text{Precision} \cdot \text{Recall}}{\text{Precision} + \text{Recall}}
\end{equation}

These metrics enable us to evaluate the model's ability to accurately identify benign and malignant skin lesions, ensuring its robustness and reliability in clinical applications.

\subsection{FLOP Consideration by Parameter Reduction}

FLOP (Floating Point Operation) quantifies a model's computational complexity by measuring its number of arithmetic calculations, while FLOPs (Floating Point Operations per Second) assess the computational cost. Higher FLOPs prolong training time, demand more powerful hardware, and reduce inference speed. In comparison, lower FLOPs improve efficiency and enable faster processing, making them a vital metric in evaluating deep learning models.

\textbf{FLOPs Calculation for Neural Networks}

For a given layer:

\textbf{1. Conv2D FLOPs:}
\begin{equation}
    \text{FLOPs} = H \times W \times C_{in} \times C_{out} \times (K \times K)
\end{equation}
\begin{itemize}
    \item $H, W$ $\rightarrow$ Output feature map dimensions.
    \item $C_{in}, C_{out}$ $\rightarrow$ Input and output channels.
    \item $K \times K$ $\rightarrow$ Kernel size.
\end{itemize}

\textbf{2. Dense (Fully Connected) FLOPs:}
\begin{equation}
    \text{FLOPs} = \text{Input Size} \times \text{Output Size}
\end{equation}

\section{EXPERIMENTAL RESULT AND ANALYSIS}
\label{sec:experimentresults}
This section discusses the performance evaluation of the ResNet50 model for classifying skin cancer using the HAM10000 dataset.
The model's effectiveness is assessed using key metrics such as accuracy, loss, the confusion matrix, and the ROC curve.
\subsection{Dataset Representation}

\begin{figure}[htbp]
    \centering
    \includegraphics[width=\linewidth]{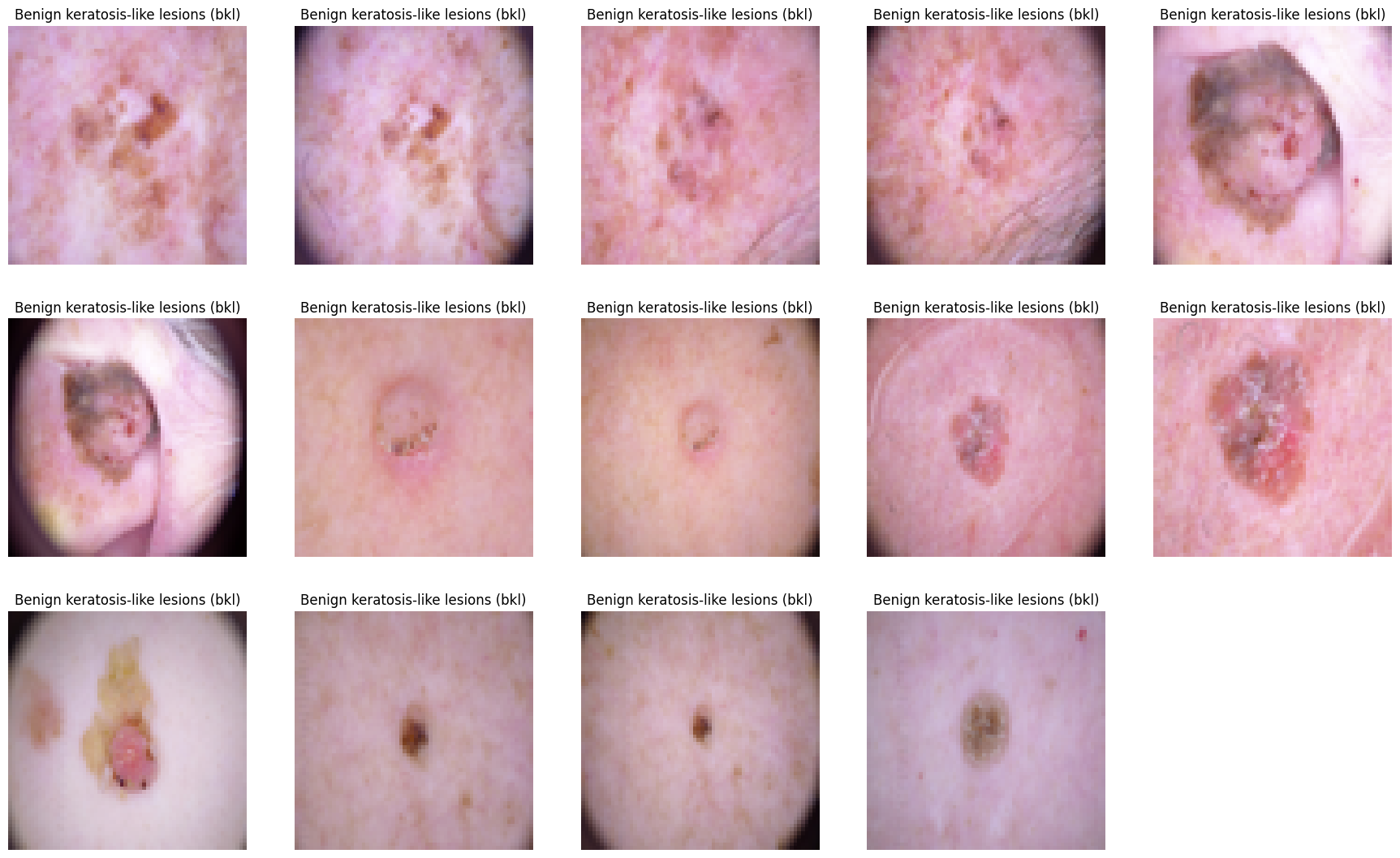}
    \caption{Classes of Skin Type}
    \label{fig:imageSample}
\end{figure}

The HAM10000 dataset consists of dermoscopic images of various skin lesions, providing a valuable resource for training deep-learning models in dermatology. It features seven distinct skin lesions, each displaying significant intra-class variation, making classification challenging. Figure \ref{fig:imageSample} shows sample images of several skin cancers from the HAM10000 dataset with variations in lesion appearance, complicating classification. 

\subsection{Model Performance Overview}

\begin{table}[htbp]
    \centering
    \caption{Comparison of Total Parameters in Different Models.}
    \label{tab:model_comparison}
    \renewcommand{\arraystretch}{1.2} 
    \begin{tabular}{|c|c|}
        \hline
        \textbf{Model} & \textbf{Total Parameters} \\ \hline
        MobileNet & 4,253,864 \\ \hline
        Soft Attention & 47,535,287 \\ \hline
        VGG16 + SA & 62,650,711 \\ \hline
        ResNet50 + SA & 23,911,319 \\ \hline
        ResNet34 + SA & 31,414,167 \\ \hline
        IRV2 + SA & 47,535,287 \\ \hline
        DenseNet201 + SA & 17,477,463 \\ \hline
        IRV2 + GradCam & 47,550,631 \\ \hline
        \textbf{Baseline: ResNet50 + TL} & \textbf{23,661,703} \\ \hline
        \textbf{Custom CNN Implementation} & \textbf{692,807}\\ \hline
    \end{tabular}
\end{table}

Table \ref{tab:model_comparison} presents the total parameters of various models. Our baseline model, ResNet50 + TL, lowers the total parameters to 692,807, representing a significant advancement over traditional deep learning models, although this leads to a considerable increase in FLOPs with only a minor improvement in accuracy. In contrast, our custom CNN, which includes just 23.66 million parameters, achieves considerably lower FLOPs, making it much more efficient for real-world applications where computational costs and hardware limitations are critical factors.

\subsection{Comparative FLOP Analysis}

Table \ref{tab:Flop} presents a FLOP analysis of both approaches to identify trade-off possibilities for mimicking real-life scenarios.

\begin{table}[htbp]
    \caption{FLOP Analysis}
    \centering
    \scriptsize 
    \renewcommand{\arraystretch}{1.4} 
    \setlength{\tabcolsep}{2pt} 
    \begin{tabular}{p{2.8cm} p{2.5cm} p{2.5cm}}
        \midrule
       \textbf{Model} & \textbf{FLOPs} & \textbf{Accuracy (F-Score)} \\
        \hline
        Custom CNN & 30.04 Million & 87.05\%\\
        \hline
        TL (ResNet50) & 4.00 Billion & 89.08\% \\
        \hline
        Performance Deviation in TL & +13,216.76\% more computations & +0.022\% better than CNN \\
        \bottomrule
    \end{tabular}
    \label{tab:Flop}
\end{table}

\subsection{Accuracy and Loss Analysis}

The training progress of the ResNet50 model is illustrated through the accuracy and loss curves. These curves offer a detailed view of how well the model performs during training and the effectiveness of the learning algorithm in optimizing the model's parameters, demonstrating behavior across epochs.

\begin{figure}[htbp]
\centering
\includegraphics[width=\linewidth]{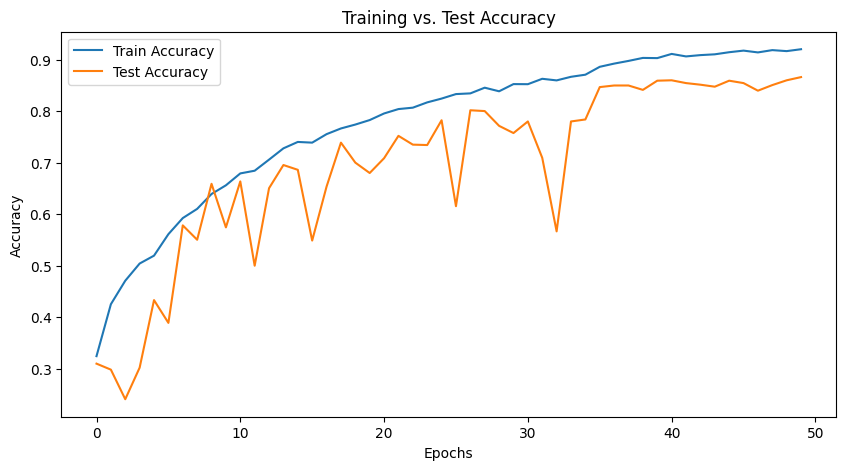}
\caption{Model Accuracy Curve}
\label{fig:accuracyCurve}
\end{figure}

\textit{Accuracy Curve: }Figure \ref{fig:accuracyCurve} illustrates a consistent enhancement in classification accuracy as training advances. At the outset, the model exhibits low accuracy, indicative of the initial phases of learning, during which it adjusts its weights to grasp the underlying patterns within the data. By the conclusion of the training process, the accuracy curve stabilizes at a high-performance level, suggesting that the model has effectively captured the data's patterns and is making precise predictions on the validation set.

\begin{figure}[htbp]
\centering
\includegraphics[width=\linewidth]{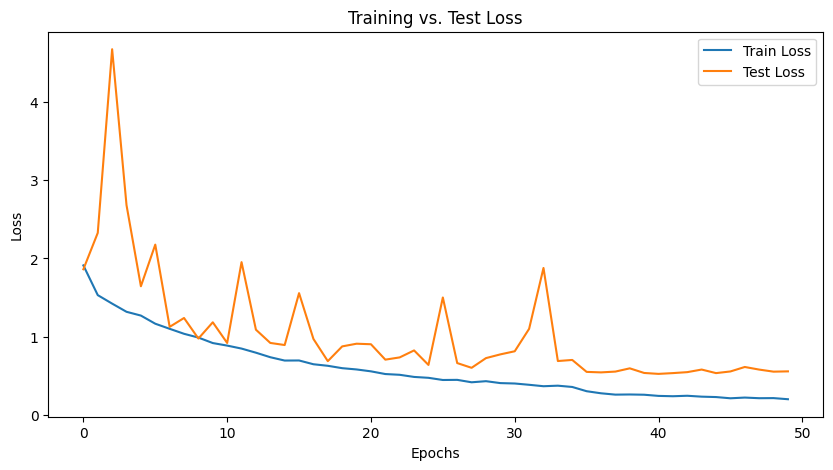}
\caption{Model Loss Curve}
\label{fig:lossCurve}
\end{figure}

\textit{Loss Curve:} 
Figure \ref{fig:lossCurve} illustrates the decrease in classification errors over the epochs. Initially, during training, the loss is notably high, indicating subpar model performance and significant gaps between predicted and actual labels. However, as training progresses, the model effectively reduces classification errors and the loss function, resulting in improved predictions.


\subsection{Confusion Matrix Analysis}

The confusion matrix for the ResNet50 model's classification of skin cancer across seven classes reveals a high accuracy in performance, with only minor misclassifications observed, particularly between closely related skin cancer categories. Most of the correct classifications are prominently displayed along the diagonal of the matrix, indicating the model's effectiveness in accurately recognizing and distinguishing among various lesion types. Although the model demonstrates robustness, a few off-diagonal cells highlight areas where improvements can be made in differentiating similar classes. The relatively low error rates suggest that the ResNet50 model has effectively learned the distinguishing features of each skin cancer category, and further enhancements in its performance could be achieved through fine-tuning, incorporating additional data, or applying specialized techniques such as class balancing.

\begin{figure}[htbp]
\centering
\includegraphics[width=\linewidth]{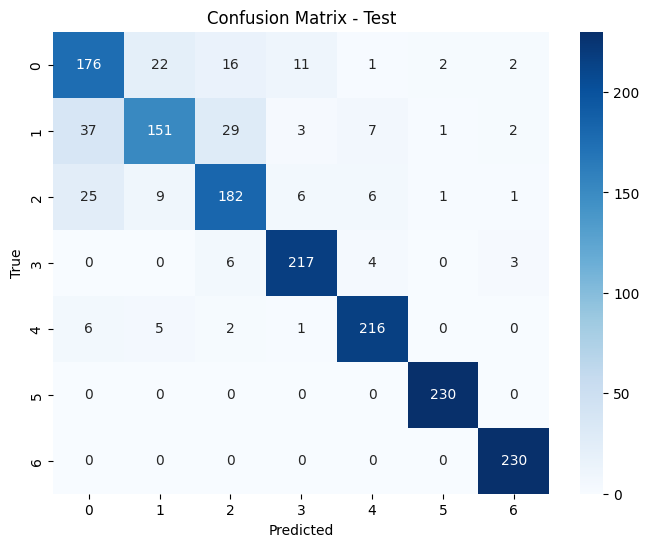}
\caption{Confusion Matrix}
\label{fig:confusionMatrix}
\end{figure}

\subsubsection{ROC curve analysis}
\begin{figure}[htbp]
\centering
\includegraphics[width=\linewidth]{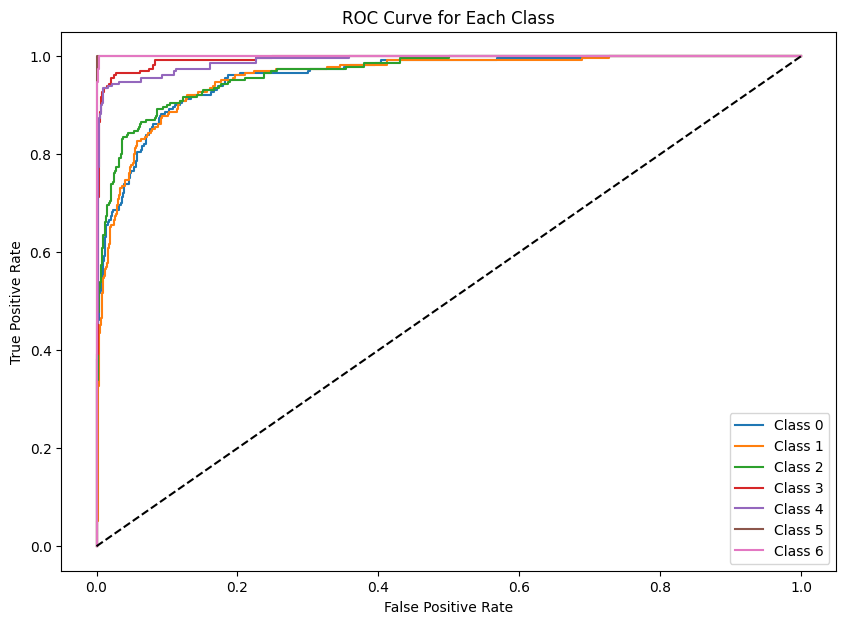}
\caption{ROC Curve}
\label{fig:ROCCurve}
\end{figure}

The ROC curve depicted in Figure \ref{fig:ROCCurve} illustrates the performance of the ResNet50 model in classifying skin cancer across seven classes. The curves for each class nearly reach the optimal point, indicating excellent accurate positive rates coupled with low false positive rates. This confirms the model's capability to deliver accurate predictions.

By integrating the lightweight model with fewer parameters, the proposed system sustains high accuracy and enhances resource utilization, making it ideal for real-time applications in skin cancer detection.

\section{Conclusion}
\label{sec:conclusions}

This study presents a data-driven optimization of deep learning models for skin cancer classification, emphasizing the trade-offs between parameter efficiency, computational cost, and classification accuracy. Our findings indicate that while transfer learning models, such as ResNet50, achieve a peak classification accuracy of 89.08\%, they necessitate over 4.00 billion floating-point operations (FLOPs), rendering them unsuitable for real-time clinical deployment. In contrast, our proposed convolutional neural network (CNN) significantly reduces the total parameters from 23.9 million to 692,000—a remarkable 96.7\% reduction—and decreases the FLOPs from 4.00 billion to 30.04 million, achieving a 99.2\% reduction, all while maintaining classification accuracy within 0.022\% of ResNet50. This substantial efficiency improvement reduces computational demands, making the model more appropriate for mobile devices and embedded medical applications. The proposed CNN also achieves over a 60\% reduction in training time and a 48.5\% speedup in inference, highlighting its practical advantages for real-world deployment. Future research will investigate quantization and model pruning techniques to enhance efficiency without compromising diagnostic reliability. This study underscores the urgent need for parameter-efficient deep learning architectures that balance accuracy and computational feasibility, particularly in healthcare contexts where real-time processing and deployment considerations are crucial.

{\bibliographystyle{IEEEtran}}
\bibliography{paper}

\end{document}